\documentclass[
reprint,
superscriptaddress,
amsmath,amssymb,
prb,
]{revtex4-2}

\usepackage{graphicx}
\usepackage{dcolumn}
\usepackage{bm}
\usepackage{tabularx}

\usepackage{soul} 
\usepackage[normalem]{ulem}

\usepackage{xcolor}
\usepackage[colorlinks=true, allcolors=blue!85!black]{hyperref}
\sethlcolor{yellow}
\usepackage{siunitx}

\usepackage{hyperref}

\usepackage[mathlines]{lineno}

\begin{document}

\preprint{APL}

\title{Dynamics of confined acoustic phonon modes in nanocrystalline silicon membranes}

\author{Tiago E. C. Magalhães}
\altaffiliation{Corresponding authors: tiago.magalhaes@inl.int, clivia.sotomayor@inl.int}
\affiliation{ 
INL - International Iberian Nanotechnology Laboratory, Av. Mestre José Veiga s/n, 4715-330, Braga, Portugal
}%

\author{Tânia M. Ribeiro}%
\affiliation{ 
INL - International Iberian Nanotechnology Laboratory, Av. Mestre José Veiga s/n, 4715-330, Braga, Portugal
}%

\author{Oili M. E. Ylivaara}%
\affiliation{VTT Technical Research Centre of Finland
Ltd., FI-02044 Espoo, Finland}%

\author{Jouni Ahopelto}%
\affiliation{VTT Technical Research Centre of Finland
Ltd., FI-02044 Espoo, Finland}%

\author{Clivia M. Sotomayor Torres}
\altaffiliation{Corresponding authors: tiago.magalhaes@inl.int, clivia.sotomayor@inl.int}
\affiliation{ 
INL - International Iberian Nanotechnology Laboratory, Av. Mestre José Veiga s/n, 4715-330, Braga, Portugal
}%

\date{\today}

\begin{abstract}

We report femtosecond time-resolved measurements of confined acoustic phonons in free-standing nanocrystalline silicon membranes and compare them with the crystalline silicon counterpart. While the latter exhibits well-resolved higher-order modes, a strong suppression of these modes is observed in nanocrystalline samples. The suppression is strongly frequency dependent and becomes more pronounced at shorter phonon wavelengths, consistent with stronger scattering by the nanocrystalline microstructure. Analysis of the phonon lifetimes reveals an additional frequency-dependent scattering mechanism in nanocrystalline silicon, with a scattering rate that scales approximately as $f^2$, where $f$ is the phonon frequency. The measured sound velocity is consistent with previous reports for nanocrystalline silicon and indicates an effective elastic response arising from multiple crystallographic orientations. These results establish coherent phonons as a sensitive probe of microstructure-dependent scattering in nanocrystalline materials and indicate that grain boundaries act as an effective spectral filter for high-frequency acoustic phonons.

\end{abstract}


\maketitle


Nanocrystalline silicon (nc-Si) is a pivotal material for nano-optoelectronics and optomechanical devices, exhibiting optical, mechanical and thermal properties that differ from those of crystalline silicon due to its nanoscale grain structure, reduced crystallinity, elastic inhomogeneity and the inherent presence of grain boundaries~\cite{wang2011thermal, claudio2014nanocrystalline, navarro2020properties, sharma2021nanocrystalline}. These structural features have a strong influence on the material’s optical absorption, mechanical dissipation, and phonon-mediated thermal transport~\cite{bagolini2017confinement, navarro2020properties, cao2024highly}. Grain boundaries introduce additional scattering channels for phonons, which suppress short-wavelength phonons and shift heat transport toward lower frequencies~\cite{maldovan2013narrow}. This leads to a strong reduction of thermal conductivity in nanocrystalline silicon through the suppression of phonon mean free paths and a pronounced dependence on grain size~\cite{wang2011thermal, klemens1994phonon, jugdersuren2021effect}. 
The crystallite size and associated grain-boundary network in nanocrystalline silicon can be tuned by thermal annealing, providing a platform for tailoring optical losses, mechanical dissipation, and phonon transport in nanophononic devices~\cite{navarro2018nanocrystalline, navarro2020properties, maire2022thermal}. Molecular-dynamics simulations further show that phonon transmission across silicon grain boundaries depends strongly on phonon wavelength and interface defect structure, with higher-frequency phonons experiencing stronger suppression~\cite{kimmer2007scattering,ju2012thermal,hickman2020thermal,  fujii2022structure}. These insights show the need for experimental studies that directly probe the dynamic response of nc-Si, particularly in regimes where coherent phonons, carrier relaxation, and nanoscale disorder interact. Here we present, to the best of our knowledge, the first experimental measurements of ultrafast time-resolved confined acoustic phonon modes in free-standing nanocrystalline silicon membranes, revealing a significant suppression of higher-order phonon modes compared to crystalline silicon.

The fabrication and characterization of nanocrystalline silicon membranes used are reported in a recent work~\cite{chen2026quasi}. A layer of amorphous silicon was grown by low pressure chemical vapor deposition on Si substrate with a layer of thermal $\mathrm{SiO_2}$. The amorphous layer was then transformed into nc-Si by heat treatments carried out in an $\mathrm{N_2}$ environment. By selecting annealing temperatures of $650$, $950$ and $1050\,^\circ\mathrm{C}$ the resulting nc-Si layers developed distinctly different nanostructures. As reported in Ref.~\cite{chen2026quasi}, the grain size distribution depends on the annealing conditions and ranges from a few nanometers to values approaching the membrane thickness. After crystallization, the nc‑Si layer was patterned using optical lithography and dry etching, and the underlying oxide was selectively removed to release the structures, forming suspended nc‑Si membranes. The diameter of the membranes is~$350\,\si{\micro\meter}$ and the thickness is $220\,\mathrm{nm}$.  We thus denote $\text{nc-Si}\,650$, $\text{nc-Si}\,950$ and $\text{nc-Si}\,1050$ the samples annealed at $650\,^\circ\mathrm{C}$, $950\,^\circ\mathrm{C}$, $1050\,^\circ\mathrm{C}$, respectively. An additional single crystal silicon (c-Si) membrane was used (Norcada, Inc.) to compare the lifetimes with those of the nc-Si samples. The thickness of this reference membrane is $20\,\mathrm{nm}$ thinner than the other samples, which may reduce the measured lifetimes due to enhanced boundary scattering in ultrathin membranes~\cite{cuffe2013lifetimes}. 

\begin{figure}[t]
    \centering
    \includegraphics[width=0.99\columnwidth]{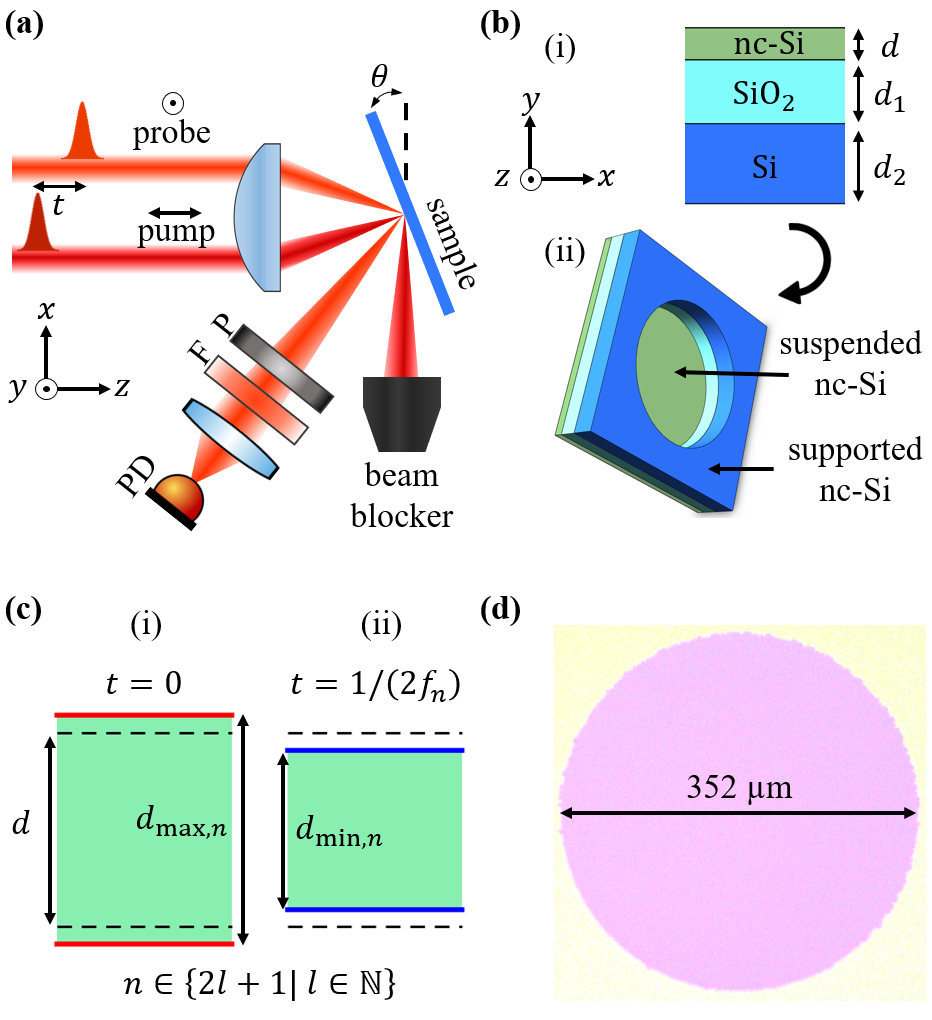}
    \caption{Experimental configuration and membrane geometry. (a) Pump-probe arrangement used to excite and detect coherent acoustic phonons in suspended nc-Si membranes. (b) (i) Cross-sectional schematic of the membrane stack (ii) Three-dimensional schematic representation of the suspended membrane (not to scale). (c) Thickness modulation associated with confined odd-order longitudinal modes $n$, given by $d_n(t)=d + \Delta d_n \cos\left(2\pi f_n t \right)$, showing the (i) maximum $d_{\max,n}$ and (ii) minimum $d_{\min,n}$ thickness states. (d) Optical microscope image (reflection mode) of a suspended nc-Si membrane (sample nc-Si 650).}
    \label{fig:fig0}
\end{figure}

The ultrafast pump-probe reflectivity measurements were performed at room temperature using the asynchronous optical sampling (ASOPS) method~\cite{bartels2007ultrafast}, where two Ti:sapphire oscillators (Gigajet TWIN, Gigaoptics GmbH), emitting 50 fs pulses, operate with a repetition rate of $\approx999.7\,\mathrm{MHz}$. The difference in repetition rate was set to $\Delta f=3000\,\mathrm{Hz}$.  The central wavelengths of the pump and probe beams are $760\,\mathrm{nm}$ and $830\,\mathrm{nm}$, respectively. The beams were focused onto each membrane at an incidence angle $\theta\approx\ang{28}$ as illustrated in Fig.~\ref{fig:fig0}. The pump and probe beams were cross-polarized to efficiently remove pump scattering at the detector. An additional long-pass filter at $800\,\mathrm{nm}$ was used to further suppress scattering from the pump beam. Both the pump and probe spot sizes are $\sim14\,\si{\micro\meter}$, which is larger than the penetration depth. This spot size yields a fluence of $28\,\si{\micro\joule/\cm^2}$  and $2.6\,\si{\micro\joule/\cm^2}$ for the pump and probe pulses, respectively.

The femtosecond pump pulse excites the lattice, launching a coherent acoustic strain pulse that can couple efficiently to the eigenmodes of the nanostructure, leading to well-defined acoustic resonances in confined geometries~\cite{thomsen1984coherent,ruello2015physical,matsuda2015fundamentals}.The small thickness of the membrane leads to the confinement and discretization of the longitudinal acoustic modes, similarly to an optical resonator~\cite{cuffe2012phonons}. Let $z$ be the coordinate across the membrane thickness as depicted in Fig.~\ref{fig:fig0}, with the two membrane--air interfaces at $z=0$ and $z=d$. The free-surface boundary condition is defined by the vanishing normal stress, i.e.,
\begin{equation} 
\left.\sigma_{zz}\right|_{z=0,d}=0\,,
\end{equation} 
where $\sigma_{zz}$ is the normal component of the stress tensor along the thickness direction. These stress-free boundaries give rise to standing longitudinal modes with discrete frequencies
\begin{equation} 
f_n = \frac{n\,v_{\mathrm{L}}}{2d}\,, \label{eq:v} 
\end{equation} 
where $n$ is the thickness-mode order and $v_{\mathrm{L}}$ is the longitudinal acoustic velocity. Only odd modes $(n=1,3,\ldots)$ modulate the membrane thickness and therefore contribute to the reflectivity signal~\cite{hudert2009confined,he2015confined}.

The detection of the generated coherent acoustic phonons relies on the modulation of the membrane reflectivity. The longitudinal dilatational modes induce a periodic variation of the membrane thickness $d(t) = d_0 +  \Delta d(t)$, which modifies the optical phase accumulated by the probe pulse in the free-standing membrane, which acts like a Fabry-Pérot cavity.  As a result, the reflectivity varies in time following the phonon oscillation. Moreover, strain can also induce changes in the refractive index via the photoelastic effect~\cite{groenen2008inelastic}. However, under the present experimental conditions, this contribution is relatively weak compared to the modulation of the optical path length associated with the thickness oscillation since the photoelastic response scales with both the photoelastic coefficient and the strain amplitude generated by the coherent phonons.  Thus, the measured reflectivity directly tracks the coherent phonon dynamics through the time-dependent interference condition of the membrane. Figure~\ref{fig:fig0} summarizes the experimental geometry, membrane structure, and the thickness modulation associated with the confined longitudinal acoustic modes investigated in this work. Additional structural characterization of membranes from the same sample set is reported in Ref.~\cite{chen2026quasi}. 
\begin{figure}[t]
    \centering
    \includegraphics[width=0.99\columnwidth]{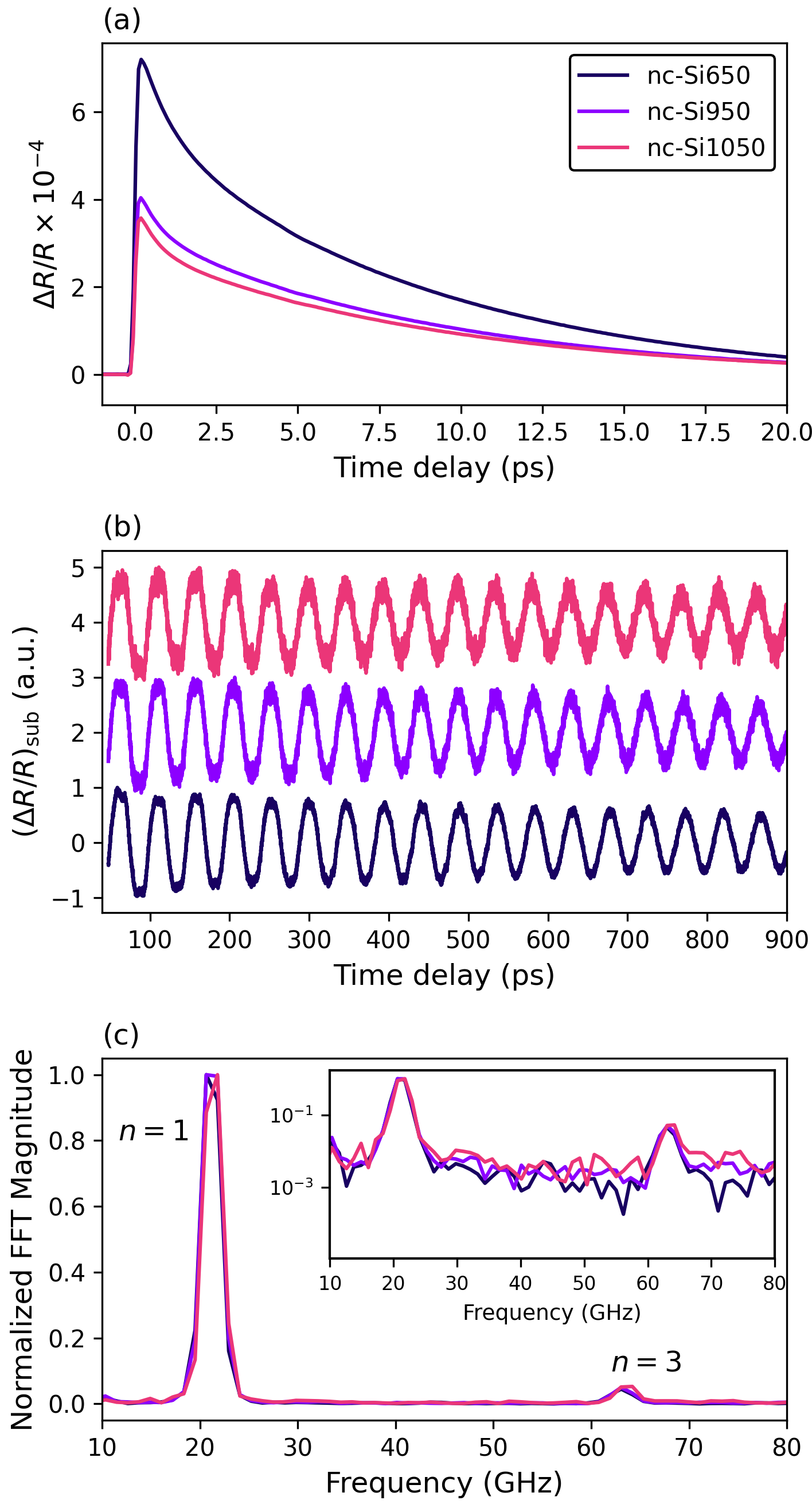}
    \caption{Ultrafast acoustic phonon modes in nanocrystalline silicon. (a) First 20 picoseconds after pump excitation. (b) Coherent phonon oscillations after multiexponential fitting subtraction. (c) Magnitude of the FFT, normalized to the first mode, revealing the first and third modes. The inset shows the data in logarithmic scale.
    }
    \label{fig:transient_data}
\end{figure}

The first picoseconds of the ultrafast time-resolved measurements in the nanocrystalline samples are shown in Fig.~\ref{fig:transient_data}(a). A multiexponential fitting is applied to the data from each sample using:
\begin{equation}
    \frac{\Delta R}{R} = \sum_{i=1}^{N} A_i\,\mathrm{e}^{- \frac{t-t_0}{\tau_{e,i}}}\left[1-\mathrm{erf}\left(\frac{w}{2\tau_{e,i}}-\frac{t-t_0}{w} \right) \right]\,,\label{eq:main}
\end{equation}
where $\tau_{e,i}\,(i=1,...,N)$ denotes the electronic relaxation times,  $A_i$ are amplitude coefficients, and $w$ is related to the setup temporal resolution, corresponding to the standard deviation of the cross-correlation of two Gaussian functions~\cite{ribeiro2025observation}. Due to high repetition-rate accumulation effects, more than two decay components are required to describe the transient response, and an optimal fit is obtained using $N=5$. The shortest decay component extracted from Eq.~(\ref{eq:main}), $\tau_{e,1}$, is of the order of 1 ps and decreases with increasing annealing temperature, taking values of $1.12\,\mathrm{ps}$, $0.74\,\mathrm{ps}$, and $0.65\,\mathrm{ps}$ for nc-Si 650, nc-Si 950, and nc-Si 1050, respectively, with the corresponding amplitudes remaining comparable among the samples. These values should be regarded as effective decay times within the temporal resolution of the experiment. The average fitted value of the temporal parameter $w$ in Eq.~(\ref{eq:main}) is $w= 95.02\pm0.66\,\mathrm{fs}$, corresponding to a temporal resolution of approximately $160\,\mathrm{fs}$. Thus, faster unresolved relaxation processes may also contribute to the transient response. A detailed investigation of the electronic relaxation dynamics is beyond the scope of this Letter and is left for future work.
\begin{figure}[t]
    \centering
\includegraphics[width=0.99\columnwidth]{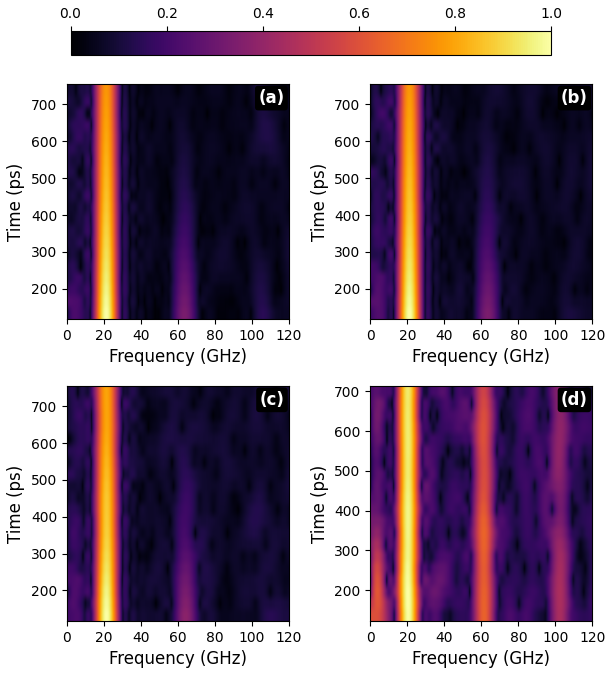}
    \caption{Short-time Fourier transform of the background-subtracted reflectivity signal $\Delta R/R_{\mathrm{sub}}$. The color scale represents the normalized square root of the magnitude $|S(f,t)|$ of the short-time Fourier transform. (a) nc-Si 650, (b) nc-Si 950, (c) nc-Si 1050, and (d) c-Si. The fundamental ($n=1$), third and fifth modes are clearly resolved in the crystalline sample, while a strong suppression of the higher-order mode is observed in the nanocrystalline membranes, where the fifth mode is hardly observed.}
    \label{fig:stft}
\end{figure}
\begin{figure}[t!]
    \centering
    \includegraphics[width=0.99\columnwidth]{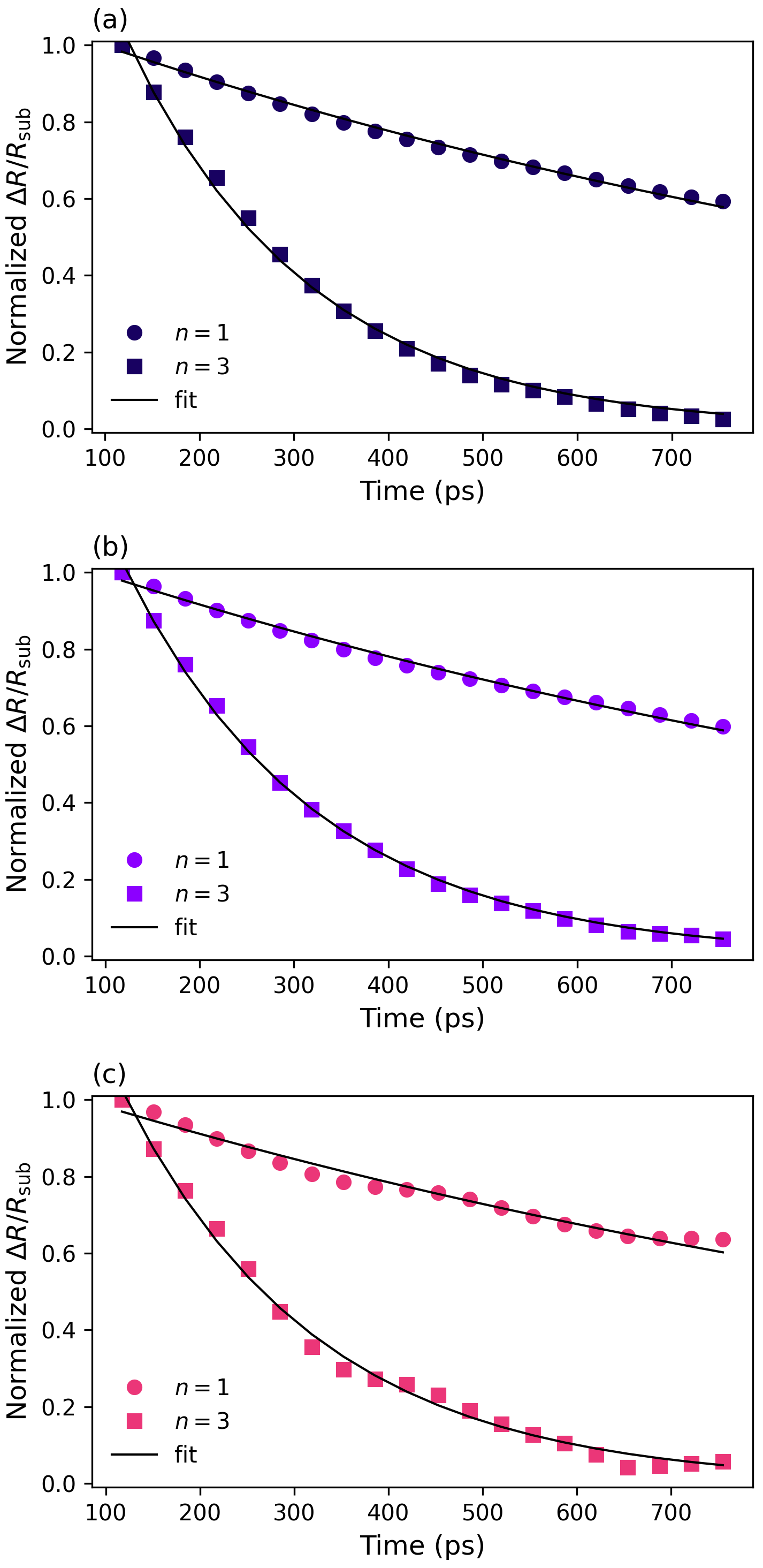}
    \caption{Decay curves with fitting for modes $n=1$ and $n=3$ for the three nc-Si samples. (a) nc-Si 650. (b) nc-Si 950. (c) nc-Si 1050.
    }
    \label{fig:fig2}
\end{figure}

The fitted background is subtracted from the experimental data, yielding the signal $\Delta R/R_{\mathrm{sub}}$ associated with the mechanical oscillations, as shown in Fig.~\ref{fig:transient_data}(b). The data is then Fourier transformed and normalized with respect to the first mode. The resulting spectrum is shown in Fig.~\ref{fig:transient_data}(c), where the first ($n=1$) and third ($n=3$) modes are clearly observed. The decrease in amplitude with increasing mode order reflects both the mode-dependent excitation by the pump-induced stress and the optical detection sensitivity to the resulting thickness and refractive-index modulations~\cite{hudert2009confined}. In the nc-Si membranes, the stronger frequency-dependent damping further suppresses the higher-order modes, causing them to approach the experimental noise level. Using the mode frequencies, $n=1$ and $n=3$, we extract the longitudinal acoustic velocity $v_\mathrm{L}$ for each sample using Eq.~(\ref{eq:v}) (see Table~\ref{tab:table1_transposed}), where $f_n$ is obtained directly from the FFT-derived resonance frequencies. The estimated $v_\mathrm{L}$ values for the nc-Si membranes are higher than that of the c-Si reference and fall close to the longitudinal sound velocity of crystalline Si along the [111] direction ($\sim9300\,\mathrm{m/s}$). This observation is consistent with previous measurements on nanocrystalline silicon films. In these studies, the effective sound velocity was found to lie between the values of single-crystal Si along different crystallographic directions and was attributed to the polycrystalline nature of the material and the averaging of elastic properties over multiple grain orientations \cite{navarro2020properties,chen2026quasi}.

The oscillation signal $\Delta R/R_{\mathrm{sub}}$ for each sample was used to perform a short-time Fourier transform (STFT) with a Hann window~\cite{oppenheim1999discrete, hofmann2013intrinsic}. The STFT was adopted to isolate the temporal evolution of the individual harmonics, verify their spectral stability throughout the measurement window, and reduce parameter correlations that arise when several damped oscillations are fitted simultaneously. The window size was chosen to be $\Delta t = 5/f_{1}$, corresponding to approximately five oscillation periods of the fundamental mode, to obtain a balance between time and frequency resolution. The resulting maps are shown in Fig.~\ref{fig:stft} for all samples. The square root of the normalized short-time Fourier transform magnitude is displayed to enhance weaker spectral components. Although the fifth harmonic is observable in the crystalline membrane and weakly present in sample nc-Si 650, its amplitude in the remaining nanocrystalline samples is comparable to the noise level. This results in substantially larger uncertainties in the extracted lifetimes. We therefore focused the analysis on the first and third modes, which could be consistently resolved across the full set of samples. We thus selected the frequencies corresponding to the first and third modes and performed a single exponential fitting, which is shown in Fig.~\ref{fig:fig2} for the three nc-Si samples. As expected, the lifetime decreases with increasing frequency. The effective phonon mean free path, $\Lambda(f)$, for each mode can be directly calculated with the measured lifetimes, $\tau(f)$, using $\Lambda(f) = v_{\mathrm{L}} \, \tau(f)$. Note that $\tau(f)$ characterizes the decay of the coherent membrane mode rather than the intrinsic lifetime of an individual acoustic phonon. Thus, $\Lambda(f)$ should be interpreted as an effective propagation length that includes all mechanisms contributing to the measured mode decay. The additional decay channel introduced by grain boundaries is expected to be responsible for the reduction of the effective mean free path compared to c-Si, which becomes more pronounced at higher frequencies.

To model the results of the nanocrystalline membranes, we use Matthiessen’s rule, i.e., we assume independent scattering mechanisms that contribute additively to the total decay rate. The total phonon decay rate can be written as \cite{klemens1994phonon, ziman2001electrons, cravero2024glass}
\begin{equation}
\tau^{-1}(f) =\tau^{-1}_{\text{int}}(f) + \tau^{-1}_{\mathrm{gb}}(f)\,,
\end{equation}
where $\tau_{\text{int}}$ describes intrinsic anharmonic phonon--phonon scattering together with boundary scattering at the membrane surfaces, and $\tau_{\mathrm{gb}}(f)$ represents additional scattering arising from grain boundaries and other microstructural effects, such as local stress and disorder, in nanocrystalline silicon. Note that this decomposition is approximate, as it neglects possible interdependence among different scattering processes, which can introduce additional scattering channels beyond the simple sum of individual rates~\cite{feng2015coupling}, so that the extracted contributions should be regarded as effective.

In general, the intrinsic contribution may include both anharmonic phonon--phonon scattering and boundary scattering, each with its own characteristic frequency dependence. In practice, separating these contributions experimentally is challenging, and they are often treated together as an effective ``intrinsic'' response~\cite{cuffe2013lifetimes,wang2011thermal}. Thus, rather than parameterizing these contributions explicitly, the intrinsic decay is obtained directly from the crystalline reference membrane and used as a baseline for the nanocrystalline samples. We used the following power-law dependence to fit the intrinsic lifetime $\tau_{\mathrm{int}}$
\begin{equation}
\tau_{\mathrm{int}}(f) = \tau_{0}\,f^{-m}\,, \label{eq:intrinsic}
\end{equation}
where $\tau_{0}$ is a prefactor and $m$ is the scaling exponent. Although the c-Si membrane has a thickness of $200\,\mathrm{nm}$, compared with $220\,\mathrm{nm}$ for the nc-Si samples, the associated variation in lifetime due to this 20 nm difference is expected to be small when compared with the much larger lifetime differences observed between c-Si and nc-Si. The c-Si membrane can therefore be used as a baseline for the analysis of the nanocrystalline membranes. Moreover, the intrinsic decay is primarily governed by the phonon frequency. Thus, the intrinsic contribution from the c-Si sample can be directly evaluated at the phonon frequencies measured in each nanocrystalline membrane.

The grain boundary contribution is modeled phenomenologically by introducing a frequency-dependent scattering rate, such that the total decay rate becomes
\begin{equation}
\tau^{-1}_{\mathrm{gb}}(f) = B\,f^{\,p}\,,\label{eq:fit}
\end{equation}
where $f$ is the phonon frequency, $B$ is a prefactor describing the strength of grain-boundary scattering, and $p$ is a scaling exponent characterizing its frequency dependence. The exponent $p$ characterizes the frequency dependence of the effective scattering process and can provide insight into the underlying scattering mechanisms \cite{cuffe2013lifetimes}. Larger values of $p$ indicate a stronger increase in the scattering rate with frequency. The parameters $B$ and $p$ can be obtained directly from fits to the experimental data of each nanocrystalline sample. 
\begin{figure}[t]
    \centering
    \includegraphics[width=0.99\columnwidth]{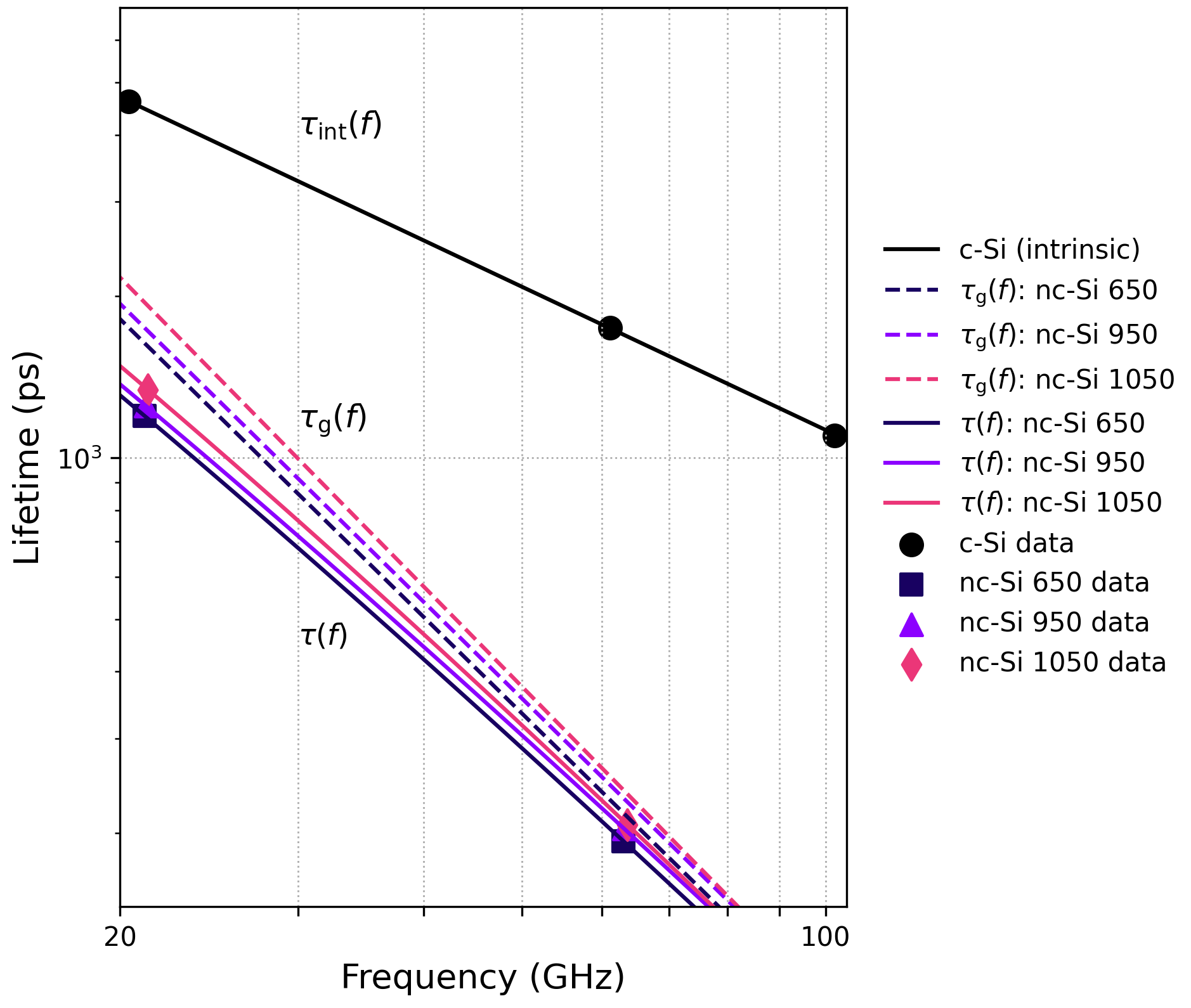}
    \caption{Phonon lifetimes as a function of frequency for crystalline and nanocrystalline silicon membranes. Symbols correspond to experimental data, while solid lines represent the total fitted model $\tau(f) = [\tau_{\mathrm{int}}^{-1}(f) + B f^{p}]^{-1}$. The black circles denote the crystalline Si reference, and the solid black line corresponds to the intrinsic lifetime $\tau_{\mathrm{int}}(f)$ given by Eq.~(\ref{eq:intrinsic}).}
    \label{fig:fig3}
\end{figure}
\begin{table*}[t]
\caption{\label{tab:table1_transposed}Summary of the extracted acoustic and transport parameters of nanocrystalline and crystalline silicon membranes. The first and third mode frequencies ($f_1$, $f_3$), longitudinal sound velocity ($v_{\mathrm{L}}$), phonon lifetimes ($\tau_1$, $\tau_3$), and corresponding mean free paths ($\Lambda_1$, $\Lambda_3$) are reported. The parameters $B$ and $p$ characterize the frequency-dependent grain-boundary scattering rates obtained from fits to Eq.~(\ref{eq:fit}). The exponent $m$ is obtained from the fit of the intrinsic lifetime of the c-Si reference using Eq.~(\ref{eq:intrinsic}). The uncertainties in $B$ and $p$ were estimated by Monte Carlo propagation of the experimental uncertainties in the measured frequencies and lifetimes through the fitting procedure.}
\begin{ruledtabular}
\begin{tabular}{lcccc}
 & nc-Si 650 & nc-Si 950 & nc-Si 1050 & c-Si\footnote{Thickness: 200 nm} \\
\hline
$f_1\,(\mathrm{GHz})$ & $21.15\pm0.21$ & $21.21\pm0.21$ & $21.31\pm0.21$ & $20.41\pm0.20$ \\
$f_3\,(\mathrm{GHz})$ & $63.09\pm0.21$ & $63.19\pm0.21$ & $63.64\pm0.21$ & $61.22\pm0.20$ \\
$v_{L}\,(\mathrm{m/s})$ & $9253\pm29$ & $9267\pm29$ & $9333\pm29$ & $8162\pm27$ \\
$\tau_1\left(\mathrm{ps}\right)$ & $1198\pm20$ & $1251\pm24$ & $1341\pm50$ & $4614\pm20$ \\
$\tau_3\left(\mathrm{ps}\right)$ & $193.2\pm3.9$ & $203.9\pm2.8$ & $207.6\pm4.7$ & $1748.4\pm5.6$ \\
$\Lambda_1\,(\si{\micro\meter})$ & $11.08\pm0.19$  & $11.60\pm0.22$  & $12.51\pm0.47$ & $37.66\pm0.20$  \\
$\Lambda_3\,(\si{\micro\meter})$ & $1.788\pm0.037$  & $1.890\pm0.024$  & $1.937\pm0.045$ & $14.271\pm0.066$  \\
$p$ & $1.844\pm0.036$ & $1.846\pm0.034$ & $1.913\pm0.059$ & $-$ \\
$B$ ($10^{6}\,\mathrm{s}^{-1}\mathrm{GHz}^{-p}$)& $2.20\pm0.30$ & $2.05\pm0.29$ & $1.49\pm0.36$ & $-$ \\
$m$ & $-$ & $-$ & $-$ & $0.8864\pm0.0035$ \\
\end{tabular}
\end{ruledtabular}
\end{table*}

We extracted the grain boundary contribution independently for each nanocrystalline sample by combining the intrinsic decay obtained from the crystalline reference with the experimental data and using Eq.~(\ref{eq:fit}). Figure~\ref{fig:fig3} shows the frequency dependence of the phonon lifetimes for crystalline and nanocrystalline membranes, together with the corresponding model fits. The obtained scaling exponents for nc-Si 650, nc-Si 950 and nc-Si 1050 were, respectively, $1.844$, $1.846$, and $1.913$. The close agreement of the exponent across all samples indicates that the dominant scattering mechanism is the same despite differences in microstructure. A quadratic frequency dependence has previously been derived for phonon scattering by disordered intergrain regions in the low-frequency limit~\cite{klemens1994phonon}. The similarity with the present exponents provides a possible interpretation of the measured frequency dependence, although the thickness and elastic properties of the grain-boundary regions are not independently known.

In contrast, the prefactor of the grain-boundary term decreases with increasing annealing temperature, taking values of $2.20\times10^{6}$, $2.05\times10^{6}$, and $1.49\times10^{6}$ (in units of $\mathrm{s}^{-1}\mathrm{GHz}^{-p}$) for nc-Si 650, nc-Si 950, and nc-Si 1050, respectively. This reduction indicates a weaker overall scattering strength for higher annealing temperatures. Such a trend is consistent with the evolution of the nanostructure towards larger effective grain sizes and a lower volume of grain boundaries upon annealing, as commonly reported for nanocrystalline silicon~\cite{wang2011thermal}. Table~\ref{tab:table1_transposed} summarizes the main results.

Together, these results demonstrate that nanocrystalline silicon introduces an additional, frequency-dependent scattering channel that selectively suppresses higher-frequency (short-wavelength) phonons, while the underlying frequency scaling remains unchanged. This behavior is consistent with the well-established reduction of thermal conductivity in nanocrystalline silicon due to enhanced grain-boundary scattering~\cite{wang2011thermal}, and provides a direct link between phonon lifetime measurements and heat transport in materials with grain boundaries. 
Within the relaxation time approximation, the thermal conductivity can be expressed as
\begin{equation}
\kappa = \int C(f)\,v^{2}(f)\,\tau(f)\,df\,,
\end{equation}
where $C(f)$, $v(f)$, and $\tau(f)$ are the spectral specific heat, group velocity, and phonon lifetime, respectively. The strong reduction of the lifetime at higher frequencies observed here therefore implies a suppression of the corresponding contribution to thermal transport, consistent with the reduced thermal conductivity reported for nanocrystalline silicon~\cite{wang2011thermal, maldovan2013narrow}. In this context, the selective suppression of short-wavelength phonons may also influence mechanical dissipation and quality factors in nanomechanical resonators based on nanocrystalline silicon~\cite{maire2022thermal}.

In summary, we have performed the first ultrafast pump-probe measurements of confined longitudinal acoustic phonon modes in free-standing nanocrystalline silicon membranes, where a suppression of higher modes is observed. Higher-order modes are strongly suppressed in nc-Si compared to c-Si, with the fifth mode being almost undetectable, which we interpret as efficient scattering of short-wavelength coherent phonons by grain boundaries and elastic heterogeneity. Our results show that nc-Si filters high-frequency acoustic modes, which could be important for engineering phonon spectra and reducing thermal noise in nanoscale optomechanical and phononic devices.

\begin{acknowledgments}
T.E.C.M., O.M.E.Y., J. A. and C.M.S.T. acknowledge support from the ERC project LEIT (grant agreement 885689). T.M.R., O.M.E.Y., J. A., and C.M.S.T. acknowledge support from the MAGNIFIC project (HORIZON-CL4-2022-RESILIENCE-01, Grant Agreement No. 101091968).
\end{acknowledgments}

\bibliography{mybib}

\end{document}